\documentclass[pre,aps,twocolumn]{revtex4-1}

\usepackage[pdftex]{graphicx}
\usepackage{amsbsy,amssymb,amsmath,bm,mathtools}

\usepackage{xspace}
\usepackage{bm}
\usepackage{color}
 
\usepackage{url}
\usepackage[section]{placeins}
\usepackage[colorlinks=true,linkcolor=blue,citecolor=blue]{hyperref}

\begin{document}


\title{Machine learning for structure-property relationships: Scalability and limitations}


\author{Zhongzheng Tian}
\author{Sheng Zhang}
\author{Gia-Wei Chern}
\affiliation{Department of Physics, University of Virginia, Charlottesville, VA 22904, USA}

\date{\today}

\begin{abstract}
We present a scalable machine learning (ML) framework for predicting intensive properties and particularly classifying phases of many-body systems. Scalability and transferability are central to the unprecedented computational efficiency of ML methods.  In general, linear-scaling computation can be achieved through the divide and conquer approach, and the locality of physical properties is key to partitioning the system into sub-domains that can be solved separately. Based on the locality assumption, ML model is developed for the prediction of intensive properties of a finite-size block. Predictions of large-scale systems can then be obtained by averaging results of the ML model from randomly sampled blocks of the system. We show that the applicability of this approach depends on whether the block-size of the ML model is greater than the characteristic length scale of the system. In particular, in the case of phase identification across a critical point, the accuracy of the ML prediction is limited by the diverging correlation length.  The two-dimensional Ising model is used to demonstrate the proposed framework. We obtain an intriguing scaling relation between the prediction accuracy and the ratio of ML block size over the spin-spin correlation length. Implications for practical applications are also discussed. 
\end{abstract}

\maketitle

\section{Introduction}

\label{sec:intro}

Machine learning (ML) is a fast advancing field that has reshaped many industries. ML has achieved surprising success in many real-world problems including machine vision, speech recognition, and natural language processing. In recent years, numerous successful ML applications in a wide range of  disciplines have also led to a paradigmatic shift in scientific research. The remarkable capability of modern ML methods to deduce complex patterns from large datasets has allowed scientists to derive connections between raw data and desired quantities, a task which previously would have been impossible. 
One of the most important application of ML in materials science is the fast and accurate prediction of material properties from the structural or configurational data~\cite{jung19,fan20,swanson19,hundi19,bapst20,hart21,abdusalamov21,zhang20x,schoenholz16}. 
Indeed, ML-based modeling of structure-property relationships is expected to open a new avenue for accelerated materials discoveries~\cite{schmidt19,ramprasad17,butler18,morgan20,saal20}.

Similar ML approaches to structure-property modeling also have enormous applications in condensed matter physics~\cite{carleo19,bedolla20}. A particularly interesting aspect of condensed matter systems is the emergence of complex symmetry-breaking phases or topological orders. The objectives of ML models are to  provide proper characterizations of such emergent ``structures", from which accurate predictions about the properties of the system can be made. Of particular interest is the classification of different phases and the identification of phase transitions of many-body systems~\cite{carrasquilla17,wang16,wetzel17,torlai16,wei17,nieuwenburg17,hu17,bohrdt19,gallo20,chng17}. For example, deep neural networks (NN) have been employed to distinguish the ordered or critical phases from the high-temperature disordered state in various classical spin systems~\cite{carrasquilla17,wang17,beach18,lakovlev18,singh19,wang20,albarracin22}.  With proper feature engineering, ML models have also been developed to capture topologically nontrivial phases or many-body localized states~\cite{zhang17,schindler17,hsu18,deng17,venderley18}.

Despite the impressive success of ML models in classifying the various many-body phases, the crucial issue of scalability, which is one of the main motivations for adopting ML approaches, has not been carefully addressed. In most studies mentioned above, the ML models, mostly implemented using deep-learning NNs, are designed for a specific system size and take the processed configuration of the whole system as the input. As a result, a new ML model has to be rebuilt and retrained for different system sizes. Besides the lack of transferability,  importantly, such approach cannot be feasibly generalized to realistic systems in the thermodynamic limit.  A scalable ML framework is thus required for both numerical and practical applications. Here the scalability refers to the capability of the framework to cope with arbitrary system size without changing the ML structures and parameters, such as the number of layers and neurons in a NN, while maintaining satisfactory performance.

One representative example of scalable ML frameworks is the Behler-Parrinello (BP) type schemes~\cite{behler07,bartok10} for the ML force field models in quantum MD simulations~\cite{li15,botu17,li17,smith17,zhang18dp,behler16,deringer19,mcgibbon17,suwa19,mueller20,noe20}. In such approaches, the total energy, which corresponds to the potential energy surface for atoms in the Born-Oppenheimer approximation, is partitioned into local contributions: $E = \sum_i \varepsilon_i$, where $\varepsilon_i$ is called the atomic energy associated with the $i$-th atom. The atomic energy is then assumed to depend predominantly on the immediate chemical environment within a length scale to be denoted as $r_{\rm BP}$.  The complex dependence of atomic energy on the local environment is to be approximated by an ML model. It is worth noting that the locality principle, or nearsightedness of electronic matter~\cite{kohn96,prodan05}, is implicitly employed in this framework. Crucially, the ML model in this scheme has a fixed size which is determined by the linear scale $r_{\rm BP}$ of the input local neighborhood and is independent of the system size to be modeled.  As a result,  the ML model can be used in systems of arbitrary large sizes through  partitioning.  The atomic forces which are crucial to MD simulations are obtained from the derivative of the total energy $\mathbf F_i = -\partial E / \partial \mathbf R_i$. Similar scalable ML frameworks have also been developed for multi-scale dynamical modeling of condensed-matter lattice models and correlated electron systems~\cite{ma19,liu22,zhang20,zhang21,zhang22b,cheng23}.

Recently, a ML framework similar in spirit to the BP scheme is proposed for efficient prediction of general extensive properties such as energy, entropy, and magnetization of arbitrarily large systems~\cite{mills19,saraseni20,mujal21,jung20}. In this approach, dubbed extensive deep-learning NN (EDNN), the system is first partitioned into partially overlapping sub-domains, also called ``tiles", of a fixed size. The length scale of the tiles is determined by the locality of the extensive quantity~$A$ of interest~\cite{mills19}. The overlap regions are introduced to partially account for the non-local effect involved in the determination of the extensive property. Importantly, as in the BP approach, a fixed-size NN model is developed to predict the extensive property $A_\alpha$ of the $\alpha$-th tile, and the extensive property of the whole system is given by the sum of contributions from each tile: $A = \sum_\alpha A_\alpha$. Again, as the NN model is of fixed size, it could be applied to arbitrarily large systems through the partitioning into tiles of the designed size. 

In this paper, we propose a scalable ML framework for identifying phases and phase transitions and, more generally, for predicting intensive properties of a many-body system. We note that the phase classification is a special case of intensive properties, which denote attributes of a system that is independent of system sizes. Since intensive attributes are spatially non-local in nature, their prediction poses great challenges for ML methodology. In our approach, a ML model is first developed to predict the intensive properties of a finite-size block, and the intensive property of the whole system is obtained by averaging over ML predictions of a number of randomly sampled blocks of the system. The ML model here is of a fixed size characterized by the size $\ell$ of the block, which plays a role similar to the length-scale $r_{\rm BP}$ in BP-type ML models. We further show that the prediction accuracy depends strongly on the ratio of the block size $\ell$ to the correlation length $\xi$ of the system, underscoring the importance of locality in scalable ML models.

The rest of the paper is organized as follows. A general framework of a scalable NN model for prediction of intensive properties is outlined in Sec.~\ref{sec:framework}. ML models are developed for the energy-density prediction and phase classification of the 2D Ising model.  A detailed study of the effect of the finite block size on the phase classification accuracy is presented in Sec.~\ref{sec:scaling}. In particular, we obtain a scaling relation relating the prediction error to the ratio of block size relative to the correlation length. A summary and implications of our results are presented in Sec.~\ref{sec:conclusion}.

\section{Scalable ML framework for intensive properties}

\label{sec:framework} 

In this section we present a scalable ML framework for predicting intensive properties of a many-body system. A particular application of this ML model is the classification of phases of a many-body system, as phase-labeling is a special intensive property of the system. Indeed, classification has long been a central subject in ML applications. Classification algorithms used in ML utilize input training data for the purpose of predicting the likelihood that a given new data will fall into one of the predetermined categories. However, scalability has not been a focus for computer scientists in traditional vision and audio-based applications of deep-learning. For example, most classification problems in image-processing, e.g. identification of certain objects or shapes in an image, are independent to the physical dimensions of an image. More specifically, the input of a ML model for image processing depends on the number of pixels of a picture, instead of its actual physical dimensions.

\begin{figure*}
\includegraphics[width=1.9\columnwidth]{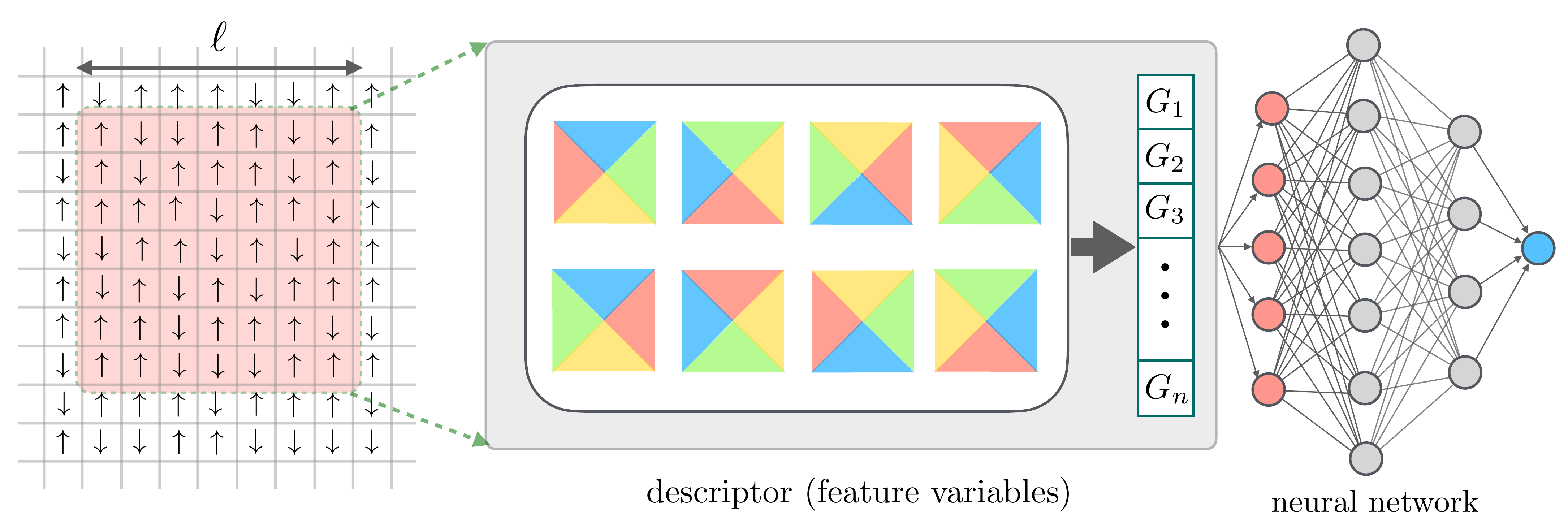}
\caption{Machine learning model for phase classification or prediction of other intensive properties of a two-dimensional Ising model. The ML model is composed of two central components: the descriptor and the neural network. The input of the ML model is a square block of Ising spins with a linear size $\ell$. The descriptor corresponds to a representation of this spin-block that is invariant with respect to symmetry operations of the $D_4$ point group of the square lattice. Essentially, the eight symmetry-related configurations are mapped to the same feature variables $\bm G = (G_1, G_2, G_3, \cdots)$, which are then fed to the input layer of the NN. The output node of the NN gives the predicted intensive property.}
\label{fig:nn-model}  \
\end{figure*}

Early applications of ML models to classification problems in condensed-matter physics also paid little attention to the issue of scalability. For example, it has been shown that a simple fully connected feed-forward NN with only one hidden layer is able to detect the two phases of the 2D Ising model and accurately reproduce the critical temperature through finite-size extrapolation~\cite{carrasquilla17}. Since each node of the input layer corresponds to an Ising spin of the 2D lattice, the size of the input layer depends on the system size. Similar ML approaches have since been developed for phase classification of numerous classical many-body systems~\cite{wang17,beach18,lakovlev18,singh19,wang20,albarracin22}. However, the architecture of such ML models obviously lacks scalability: ML model developed for a particular system size cannot be directly applied to systems of larger dimensions.

As discussed in Sec.~\ref{sec:intro}, linear scalability methods in many-body systems rely on the principle of locality, which means physical properties only depend on configurations of a finite neighborhood. This allows for a divide-and-conquer approach to the computation of the physical properties of interest. Based on this locality principle, here we discuss a scalable ML framework for efficient prediction of intensive properties and particularly phase classification of a many-body system. The limitations of this approach, due to the breakdown of locality, will be discussed in Sec.~\ref{sec:scaling}. While the formulation presented here can be applied to general many-body systems, for concreteness and as a proof of principle, we demonstrate our approach using the square-lattice ferromagnetic Ising model~\cite{2dising}: $\mathcal{H} = - J \sum_{\langle ij \rangle} \sigma_i \sigma_j$, where $\sigma_i = \pm 1$ represents an Ising spin at site-$i$, $\langle ij \rangle$ denotes a nearest-neighbor pair, and $J >0$ is the strength of the ferromagnetic interaction. This canonical spin model exhibits two thermodynamical phases: a high-temperature paramagnet with disordered spins and a long-range ordered ferromagnetic state at low temperatures. A critical point at $T_c \approx 2.269J$ separates the two phases. 

Central to our approach is the construction of a ML model which takes a finite block of Ising spins as input and predicts some intensive quantities $\mathcal{Q}$ at the output; see Fig.~\ref{fig:nn-model}. There are two central components of the ML model: the descriptor and the neural network (NN).  The spin configuration within the block is mapped to a set of feature variables $\bm G = \{ G_1, G_2, \cdots \}$, also known as a descriptor, which is invariant under the symmetry transformations of the original Hamiltonian. With the symmetrized representation $\bm G$ as the input, the NN produces the predicted intensive properties at the output nodes. The linear size $\ell$ of the spin block is a special hyperparameter of the ML model. In particular, $\ell$ determines the size of the input layer of the NN.

\subsection{Lattice descriptor} 

The construction of descriptor or feature variables is similar to the so-called image augmentation which is a common technique used to artificially expand the size and diversity of a training dataset by applying various transformations to the original images. These transformations can include rotation, scaling, flipping, translation, or color changes, among others. The goal is to make the model more robust and invariant to these transformations, which helps improve generalization and performance on new, unseen data. For applications to materials science or condensed matter physics, however, even with the augmented dataset, symmetries of the original physical Hamiltonian can only be learned approximately even with the general approximation capability of NNs. In particular, for ML models with output that are invariant under symmetry operations, a proper representation of the input variables should also be invariant with respect to the same symmetry group for consistency.  This crucial step of ML models, namely the construction of the proper descriptor, is often referred to as feature engineering~\cite{ghiringhelli15,bartok13,himanen20,rupp12,shapeev16,kondor07}.


Descriptors also play a crucial role in the scalable BP-type ML interatomic potential or force-field models in quantum MD methods. In MD applications, a proper descriptor of the atomic configuration should be invariant under rotational and permutational symmetries, while retaining the faithfulness of the Cartesian representation.  Over the past decade, several descriptors have been proposed to represent the atomic or chemical environment. Notably among them are the Coulomb matrix method~\cite{rupp12}, moment tensor potentials~\cite{shapeev16}, the atom-centered symmetry functions (ACSFs)~\cite{behler07,behler11}, and the group-theoretical bispectrum method~\cite{bartok10,bartok13}. 

For condensed matter systems defined on a lattice, the relevant symmetry operations are internal symmetries of the constituent degrees of freedom and the discrete symmetries of the lattice. Examples of the former are the $Z_2$ symmetry of Ising variables, O(2) symmetry of XY spins, and so on. For lattice systems, the symmetry of a local neighborhood is described by the point group associated with the site-symmetry. A descriptor of lattice models thus should be invariant with respect to both types of symmetry groups. A general theory of descriptors for lattice models has recently been presented in Ref.~\cite{zhang22}; several explicit implementations have also been demonstrated for well-known model systems~\cite{ma19,liu22,zhang20,zhang21,zhang21,zhang22b,cheng23}. 

For application to the Ising model, the internal symmetry group of Ising spins is $Z_2$ while the lattice symmetry is described by the $D_4$ point group. We first consider invariant representations with respect to the point-group symmetry. Essentially, the goal here is to map the eight Ising configurations related by symmetry operations of the $D_4$ group, as shown in Fig.~\ref{fig:nn-model}, to the same feature variables $\bm G = \{ G_\ell \}$. To this end, we employ group-theoretical method to obtain invariant variables of Ising configurations within a block $B_\alpha$, where $\alpha$ is an index of the block. First, we note that Ising spins $\{\sigma_j\}$ within a given block forms a high-dimensional reducible representation of the $D_4$ group, which can then be decomposed into fundamental irreducible representations (IR's) of the point group. This decomposition can be highly simplified as the representation matrix is automatically block-diagonalized, with each block corresponding to a fixed distance from the center-site of the block. We use $ {\bm  f}^\Gamma = (f^\Gamma_1, f^\Gamma_2, \cdots , f^\Gamma_{D_\Gamma})$ to denote the basis function of IR of the symmetry-type $\Gamma$. For example, four nearest-neighbor Ising spins $\{\sigma_1, \sigma_2, \sigma_3, \sigma_4\}$ form a closed representation, and can be decomposed as $4 = 1A_1 + 1B_1 + 1E$, where $f^{A_1} = \sigma_1 + \sigma_2 + \sigma_3 + \sigma_4$ and so on; more details can be found in Appendix~\ref{sec:descriptor}. 

Given these IR coefficients, one immediate class of invariants is their amplitudes $p^\Gamma = |{\bm f}^\Gamma|^2$, which is called the power spectrum of the representation. However, the descriptor also needs to account for crucial information on the relative phases of different IRs. To capture the phase information,  we introduce the concept of reference IR coefficients ${\bm f}^{\Gamma}_{\rm ref}$, which are obtained by applying similar decomposition procedure to large symmetry-related groups of Ising spins within the block $B_\alpha$ such that they are insensitive to small variations of the neighborhood~\cite{zhang22}. Importantly, the relative phase of two IRs can be restored from their respective relative phases $\eta^{\Gamma} \sim {\bm f}^{\Gamma} \cdot {\bm f}^{\Gamma}_{\rm ref}$ to the reference IR.  A complete set of invariant feature variables is then given by the power spectrum $p^\Gamma$ and the phases $\eta^\Gamma$; see Appendix~\ref{sec:descriptor} for more details. Finally, we note that the power-spectrum obviously is invariant under the $Z_2$ transformation $\sigma \to -\sigma$ and ${\bm f} \to - {\bm f}$. The relative phases also remain the same under the $Z_2$ symmetry as both the IR and the corresponding reference IR change sign. The $Z_2$ symmetry is automatically preserved in this descriptor.

\subsection{Neural network}

Given the invariant representation of the spin block $B_\alpha$, the corresponding intensive property, let's call it $\mathcal{Q}$, is assumed to depend on the feature variables through a universal function. Specifically, let $B_\alpha$ be the spin configurations within the $\alpha$-th block, the intensive property $\mathcal{Q}$ of the block is given by
\begin{eqnarray}
	\mathcal{Q}_\alpha = \mathcal{Q}(B_\alpha) = \hat{f}(\bm G_\alpha)
\end{eqnarray}
where $\bm G_\alpha = \{G_1(B_\alpha), G_2(B_\alpha), \cdots \}$ denote feature variables built from the $\alpha$-th block, and the ``universal" function $\hat{f}(\cdot)$ (universal in the sense of a given Hamiltonian) is to be approximated by a deep neural network. Because of the discrete nature of Ising spins, we include a convolutional neural network (CNN) to better represent the discretized input feature variables. {CNN is a class of neural networks that utilize the shared-weight architecture of convolutional filters to process input features. One of the main advantages of CNN is that they can effectively handle discrete data by smoothly sliding the filters over the input, which leads to the emergence of translational equivariant responses. In other words, the CNN's ability to extract local features across multiple spatial positions allows it to recognize patterns in the input data regardless of their location, making it a powerful tool for image and signal processing tasks.} This property enables CNN to capture the structural characters of the images and other 2D/3D objects. 

The output of the CNN is then fed into a fully connected feed-forward NN, which performs a sequence of transformations. Specifically, for the $m$-th layer, the neuron processes the activation $\mathbf{a}^{(m-1)}$ from the previous $(m-1)$-th layer through weight and bias parameters: {$\mathbf a^{(m)} = {\bf A}(\mathbf{w}^{(m-1)}\mathbf{a}^{(m-1)}+\mathbf{b}^{(m-1)})$, where $\bf A$ is an array of identical activation functions, $\mathbf{w}^{(m-1)}$ is the weight of the $(m-1)$-th layer, and $\mathbf{b}^{(m-1)}$ is the bias of the $(m-1)$-th layer. Finally, the node at the end of the feed-forward NN} is the predicted intensive property $\mathcal{Q}_\alpha$. These NN parameters $\mathbf w$ and $\mathbf b$ {of both CNN and feed-forward NN} are essentially fitting parameters for the high-dimensional function $f_{\rm ML}$, which are determined through training {with stochastic gradient descent.}

\begin{figure*}
\includegraphics[width=1.9\columnwidth]{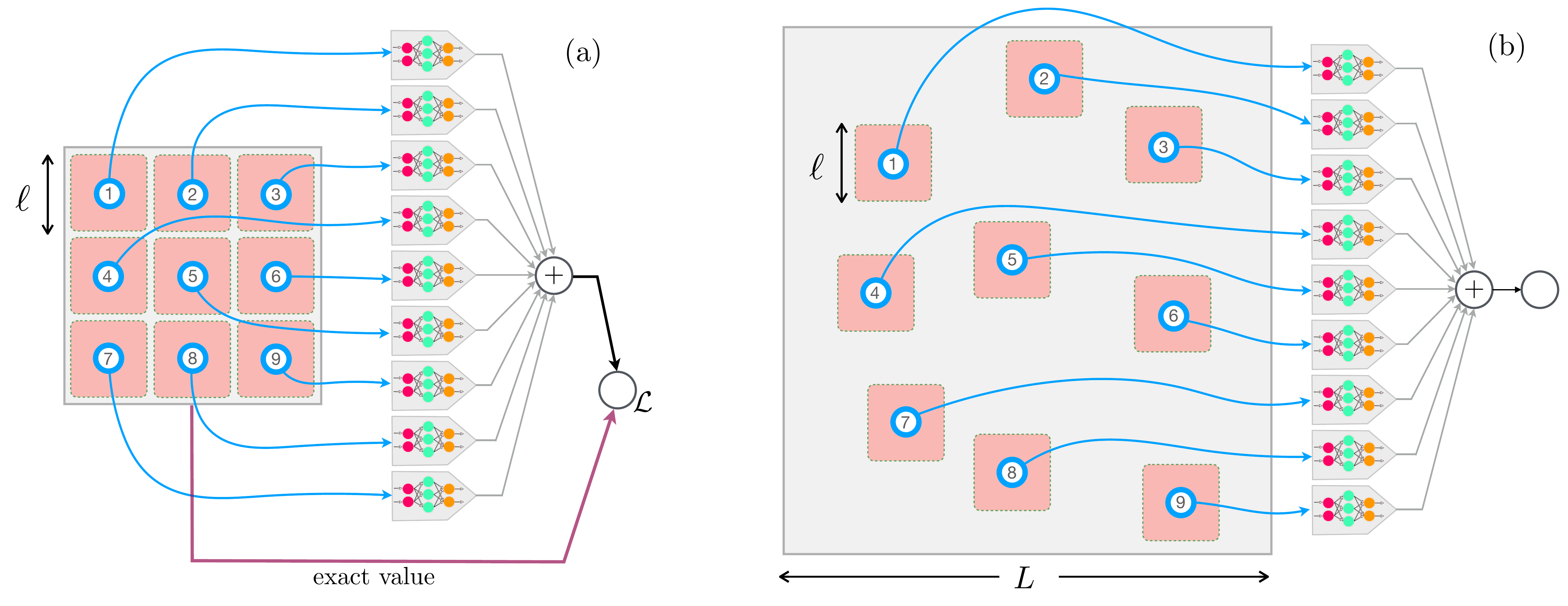}
\caption{A scalable ML approach relies on the partitioning of the system into finite-size blocks that can be solved individually. The presumably time-consuming calculation of some physical properties of the block of linear size $\ell$ is encoded in the ML model, which is implemented using the multi-layer NN here. Panel~(a) shows a schematic of the training process. The loss function $\mathcal{L}$ quantifies the difference between the exact value and the average of the ML predictions from each block.  For application of the ML framework to large systems, one can design the partitioning such that the blocks cover the whole system. However, for extremely large system $L \gg \ell$, or experimental data where the system approaches the thermodynamic limit, a practical approach is to randomly select a large number of blocks to represent the system, as shown in panel~(b). The estimation of the physical property is again the average of ML prediction from all blocks. }
\label{fig:ml-scaling}  \
\end{figure*}

\subsection{Training of the ML model}

The NN model can be optimized through standard supervised training algorithms. For example, for a spin block described by feature variables $\bm G$,  the loss function can be defined as the mean square error (MSE) $\mathcal{L} =  |\hat{f}(\bm G) - f_{\rm ML}(\bm G)|^2$. In this approach, each spin block is treated as independent training dataset. However, for highly inhomogeneous systems, large fluctuations of block-spin configuration are expected. Such situations occur, for example, in spin systems close to a critical point. In order to better account for local variations of spin-blocks, here we propose a training scheme shown schematically in Fig.~\ref{fig:ml-scaling}, which is similar in spirit to the BP-type ML models for quantum MD simulations. The training dataset is obtained from a system of linear size $L$ which could be much larger than the block size $\ell$. We partition the large Ising system into $M \sim (L/\ell)^2$ blocks, each is labeled by $B_\alpha$, where the index $\alpha = 1, 2, \cdots, M$; see Fig.~\ref{fig:ml-scaling}(a). Importantly, the training is based on averaged predictions from these $M$ blocks. For example, we consider the following MSE loss function for a given snapshot of the system
\begin{eqnarray}
	\label{eq:loss_func}
	\mathcal{L} =  \biggl| \hat{\mathcal{Q}} - \frac{1}{M}\sum_{\alpha=1}^M f_{\rm ML}\bigl(\bm G_\alpha \bigr) \biggr|^2,
\end{eqnarray}
where $\hat{\mathcal{Q}}$ is the exact intensive property of the whole $L\times L$ system, and $\bm G_\alpha$ is the feature variables for the $\alpha$-th block of the snapshot.  For phase classification problem where $\mathcal{Q} = 1$ or 0 corresponds to whether the system is in a specific phase or not, a cross entropy can be used for the loss function.

With either MSE or cross entropy loss functions, as demonstrated in Fig.~\ref{fig:ml-scaling}(a), one can view the training process as the optimization of a super-neural net build from $M$ copies of the same NN shown in Fig.~\ref{fig:nn-model}. Finally, standard stochastic gradient descent method with back propagation is used to optimize the ML model from multiple snapshots of the whole system.
It is worth noting that the above scheme is different from ML training based on $M$ independent blocks. With loss function determined from the averaged predictions from all blocks, the optimization of the ML model is forced to take into account the potentially diverse block-spin configurations within the same system. 

We emphasize again the similarity of our approach to BP-type ML structures discussed above. By partitioning the total energy of the system into local energies $E = \sum_{i=1}^N \varepsilon_i$, where $N$ is the number of atoms, the local energy associated with the $i$-th atom is assumed to depend on its immediate chemical environment, also denoted as $\mathcal{C}_i$, through a universal function, i.e. $\varepsilon_i = \epsilon_{\rm ML}(\mathcal{C}_i)$. This universal function $\epsilon_{\rm ML}(\cdot)$ is similarly to be approximated by a ML model. For MD applications, one is mostly interested in atomic forces, which are given by the derivatives, $\mathbf F^{\rm ML}_i = -\partial E / \partial \mathbf R_i$, and are dependent indirectly on the universal function. Importantly, the loss function for a given snapshot of atoms is based on prediction difference of total energy and individual forces, e.g. $\mathcal{L} =  \sum_{i} | \hat{\mathbf F}_i - \mathbf F^{\rm ML}_i|^2 + r | \hat{E} - \sum_i \varepsilon_i|^2$, where the $\hat{\mathbf F}_i$ and $\hat{E}$ denote the forces and total energy, respectively, obtained from, e.g. DFT calculations, and coefficient $r$ specifies the relative ratio of energy constraints compared to that of forces. The ML model $\epsilon_{\rm ML}(\cdot)$, which determines both the forces $\mathbf F^{\rm ML}_i$ and local energies, is optimized from the minimization of this loss function defined on the whole atomic system.

Once the ML model is optimized, its utilization for predicting intensive properties of a much larger system is based on a random sampling method. As shown in Fig.~\ref{fig:ml-scaling}(b), a number $N_{\rm sample}$ of $\ell\times \ell$ blocks are randomly selected from the whole system. Let $\mathcal{Q}_\alpha = f_{\rm ML}(\bm G_\alpha)$ be the ML prediction of the $\alpha$-th block, the intensive property of the system is approximated by averaging over these local predictions
\begin{eqnarray}
	\overline{\mathcal{Q}} =   \frac{1}{N_{\rm sample}} \sum_{\alpha = 1}^{N_{\rm sample}} f_{\rm ML}(\bm G_\alpha). 
\end{eqnarray}
One can also obtain an estimate of the prediction error from the standard deviation of the local predictions: 
\begin{eqnarray}
	\sigma_{\mathcal{Q}} = \left[ \frac{1}{N_{\rm sample}} \sum_{\alpha=1}^{N_{\rm sample}} \left(f_{\rm ML}(\bm G_\alpha) - \overline{\mathcal{Q}} \right)^2 \right]^{1/2}
\end{eqnarray}
For large number of sampling blocks, the prediction accuracy can be improved by the average $\overline{\mathcal{Q}}$. Moreover, the standard deviation $\sigma_{\mathcal{Q}}$ provides information of the intrinsic fluctuations of the intensive properties within the system measured based on the length scale $\ell$ associated with the ML model. 

As a demonstration, here we first apply the above approach to the prediction of the energy density $\rho_E = E / N$ of the 2D Ising model, where $E = -J\sum_{\langle ij\rangle} \sigma_i \sigma_j$ is the total energy of the system and $N = L^2$ is the total number of spins. It is worth noting that the energy calculation in short-range Ising model, or most classical short-range spin models, is linear-scaling and can be done rather efficiently. However, the computational time is still of the order of $\mathcal{O}(N)$, which will become infeasible in the thermodynamic limit. On the other hand, the time complexity of the ML method scales as $\mathcal{O}(N_{\rm sample})$, which is controlled by the number of sampling blocks. Moreover, the ML approach also offers a more efficient calculation of energy density for models with long-range interactions or quantum systems where the energy calculation requires time-consuming many-body methods.



{The construction of both the descriptor and the neural network (NN) model was implemented using the PyTorch library. To accelerate the training process of the machine learning (ML) model, multiple graphics processing units (GPUs) were employed in PyTorch. The training dataset was composed of 40,000 Ising configurations of a lattice with linear size $L=320$, while an additional 8,000 Ising configurations were used for validation. The temperature values of the training and validation datasets were evenly distributed across 40 different temperatures ranging from $T_{\rm low} = 1.2 J$ to $T_{\rm high} = 3.6 J$. The ML models were constructed using two block sizes $\ell = 32$ and $64$ as the input data.  Finally, the model is trained using the Adam optimizer~\cite{Kingma} with adaptive learning rate of 0.001 and 500 epochs.}

\begin{figure}
\includegraphics[width=1.0\columnwidth]{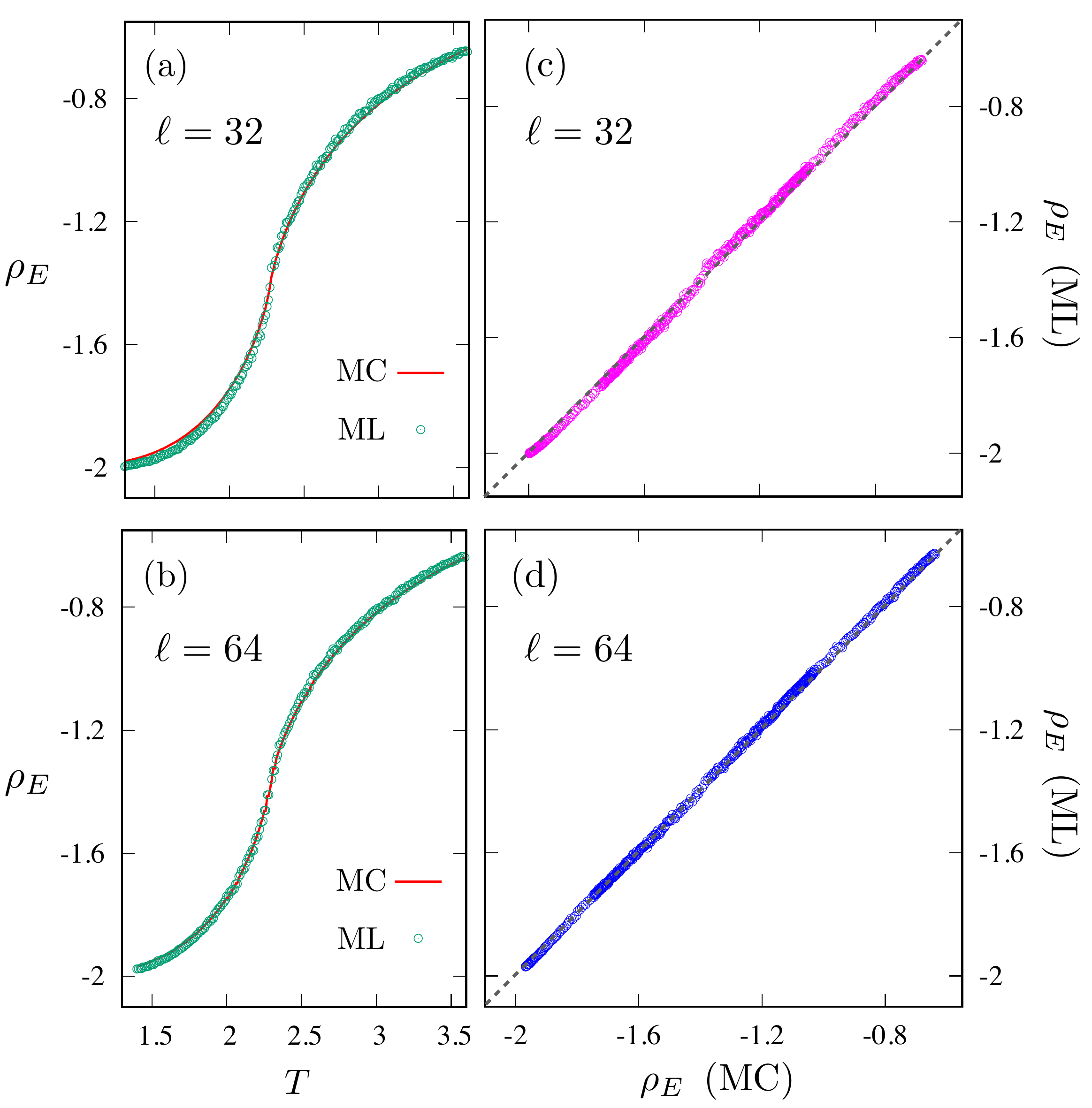}
\caption{Left panels show the energy density $\rho_E$ of the 2D Ising model as a function of temperature obtained from Monte Carlo (MC) simulations and ML predictions with block size (a) $\ell = 32$ and (b) $\ell = 64$. The right panels show the comparison of energy density predicted by ML models versus that obtained from MC simulations again for two different block sizes (c) $\ell = 32$ and (d) $\ell = 64$. }
\label{fig:e-density}  \
\end{figure}

Fig.~\ref{fig:e-density}(a) and (b) show the ML-predicted energy density versus temperature curves for an $L=320$ Ising system based on block sizes $\ell = 32$ and 64, respectively. The results from the Monte Carlo (MC) simulations are also shown for comparison. While the overall ML predictions agree well with the MC results, a systematic discrepancy can be seen from the figures:  the ML models tend to overestimate the energy density above the critical temperature and to underestimate the energy  below $T_c$; see also the comparison of ML predictions against the energy density from MC simulations in Fig.~\ref{fig:e-density}(c) and (d). This systematic errors become more noticeable for ML models with a smaller block size $\ell$. 

It should be noted that, while computing the energy of a given spin-block is almost trivial especially for short-ranged spin models, our objective here is to predict the intensive property of the whole system through a window of linear size $\ell$. This is also the source of the prediction errors observed in Fig.~\ref{fig:e-density}. The discrepancy is also due to the requirement that same ML model is trained to predict the intensive property of both high- and low-temperature phases, including the critical regime around $T_c$. Indeed, excellent prediction accuracy can be achieved if the ML model is restricted to either the high-temperature paramagnet or the long-range ordered phase at low $T$. The highly inhomogeneous spin states in the vicinity of the critical point introduces significant uncertainties in the training of finite-size ML models, giving rise to the systematic errors in both phases. Intuitively, in order to model spin configurations in the critical regime which comprises diverse spin configurations characterized by large length scales, the ML model is forced to compromise its prediction capability of both the high and low temperature phases.  A more systematic analysis of prediction error and its relation to the diverging correlation length at the critical point are discussed in the next section.

\section{Correlation Length and limitation of scalability}

\label{sec:scaling}

Although interactions of most physical systems are local and characterized by microscopic length scales of the order of lattice constants, emergent structures of the system are often described by length scales, denoted as $\xi$ for convenience, that are often much larger than the microscopic one.  More importantly, the locality of intensive quantities $\mathcal{Q}$ which depend on such emergent structures is controlled the emergent length scales.  For example, while spin-spin interactions in the standard Ising model are restricted to nearest neighbors, the spatial fluctuations of spin configurations are determined by the correlation length $\xi$ of the system. In general, the correlation length characterizes the exponential decay of the structural correlation function 
\begin{eqnarray}
	C(\mathbf r) \sim \exp(-|\mathbf r| / \xi). 
\end{eqnarray}
Importantly, the length scale $\xi$ is often not a fundamental parameter of the Hamiltonian, but is controlled by external parameters such as temperature or magnetic field.

\begin{figure}
\includegraphics[width=0.9\columnwidth]{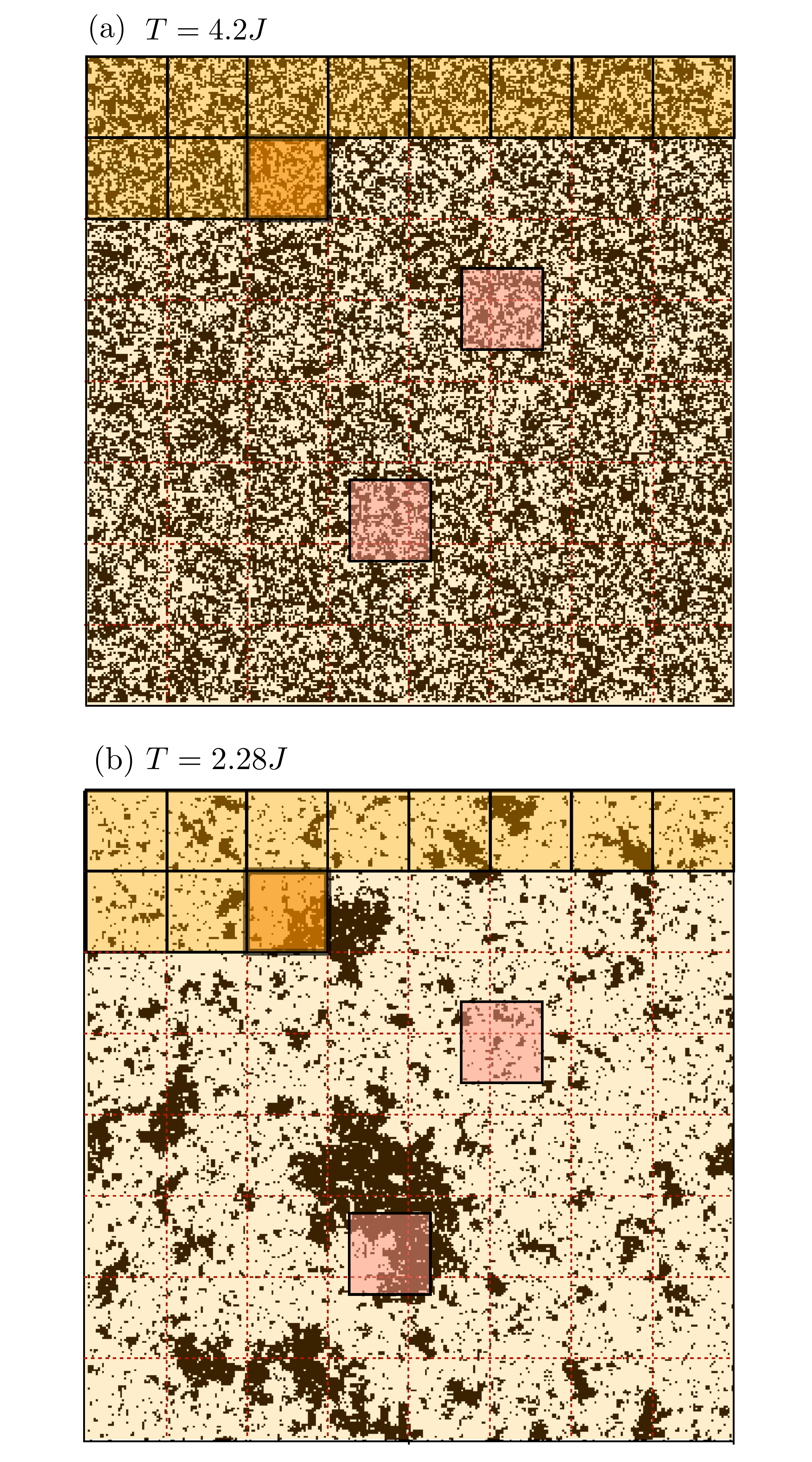}
\caption{Comparison of the block size $\ell$ and the correlation length $\xi$ of the 2D Ising model at (a) $T = 4.2J$, which is representative of the high-temperature paramagnetic phase, and (b) $T = 2.28J$ close to the critical point $T_c \approx 2.269J$.}
\label{fig:focuscoverage}  \
\end{figure}

The effects of correlation length on the performance of ML model with a finite block size $\ell$ can be intuitively understood from the two snapshots in Fig.~\ref{fig:focuscoverage}. The squares in both panels correspond to the spin-block for the input of the ML model. For systems at high temperatures, as exemplified by Fig.~\ref{fig:focuscoverage}(a), the length scales of spatial fluctuations or inhomogeneity is much smaller than the block size, $\xi \ll \ell$. Consequently, spin configurations sampled by a given block are good representatives of the whole system. As a result, ML predictions based on even one block could be accurate in this limit.

On the other hand, larger and larger spin clusters start to emerge as system approaches the critical temperature~$T_c$, as shown in Fig.~\ref{fig:focuscoverage}(b). The average linear size of spin clusters is again characterized by the correlation length $\xi$. When this length scale is greater than the block size, $\xi \gtrsim \ell$, an individual spin block is expected to exhibit rather distinct spin configurations. As a result, each block only captures partial information of the larger-scale structures of the system. By taking into account all spin blocks of the whole system in the optimization of the ML model, the training scheme outlined in Fig.~\ref{fig:ml-scaling}(a) with a loss function of Eq.~(\ref{eq:loss_func}) aims to strike a balance between spatial features of length scale $\ell$. The ML prediction shown in Fig.~\ref{fig:ml-scaling}(b) can be viewed as attempting to assemble information contained in several small pictures of size $\ell$ to form a big picture at the length scale of~$\xi$. However, due to the ambiguities and trade-off between different blocks during the training, the accuracy of ML prediction is significantly hampered in the $\xi \gg \ell$ limit.

To examine quantitatively the effects of a diverging correlation length on the ML prediction accuracy, here we develop a ML model for the phase classification of the 2D Ising model. To this end, a deep-learning NN with one non-reducing and five reducing fully connected layers was constructed.  The NN produces a single digit $\mathcal{Y}$ at the output layer such that $\mathcal{Y} = 0$ and 1 corresponds to the low-$T$ ferromagnetic order and the paramagnetic phase at high temperatures, respectively. As discussed above, the ML model is designed to provide an accurate approximation for the structure-property mapping $\mathcal{Y} = f_{\rm ML}(\bm G)$ of the spin block, where $\bm G = \{G_1, G_2, \cdots \}$ are symmetry-invariant feature variables built from spin configuration of the block discussed above. A binary cross entropy (BCE) loss function was employed for the optimization of NN model. Specifically, let $\overline{\mathcal{Y}}_m$ denotes the ML prediction averaged over all blocks of the $m$-th snapshot of an $L\times L$ Ising system, and $\hat{\mathcal{Y}}_m$ be the true value of the corresponding phase, the BCE loss function is defined as
\begin{eqnarray}
	\mathcal{L} = - \frac{1}{\mathcal{N}} \sum_{m=1}^{\mathcal{N}} \left[\hat{\mathcal{Y}}_m \log(\overline{\mathcal{Y}}_m) 
	+ (1-\hat{\mathcal{Y}}_m) \log(1 - \overline{\mathcal{Y}}_m) \right] \qquad
\end{eqnarray}
where $\mathcal{N}$ is the number of snapshots used in the training. For the phase classification, $\mathcal{N} = 50,000$ Ising configurations from Monte Carlo simulations on a $L = 320$ system were used to train the NN model, with an additional 10,000 snapshots reserved for validation. The datasets were obtained from MC simulations over a range of temperatures similar to that in the ML model for energy density. An Adam optimizer with a global learning rate of 0.001 is used. A dropout layer is added before the output layer with a dropout probability set to 0.2.

\begin{figure}
\includegraphics[width=1.0\columnwidth]{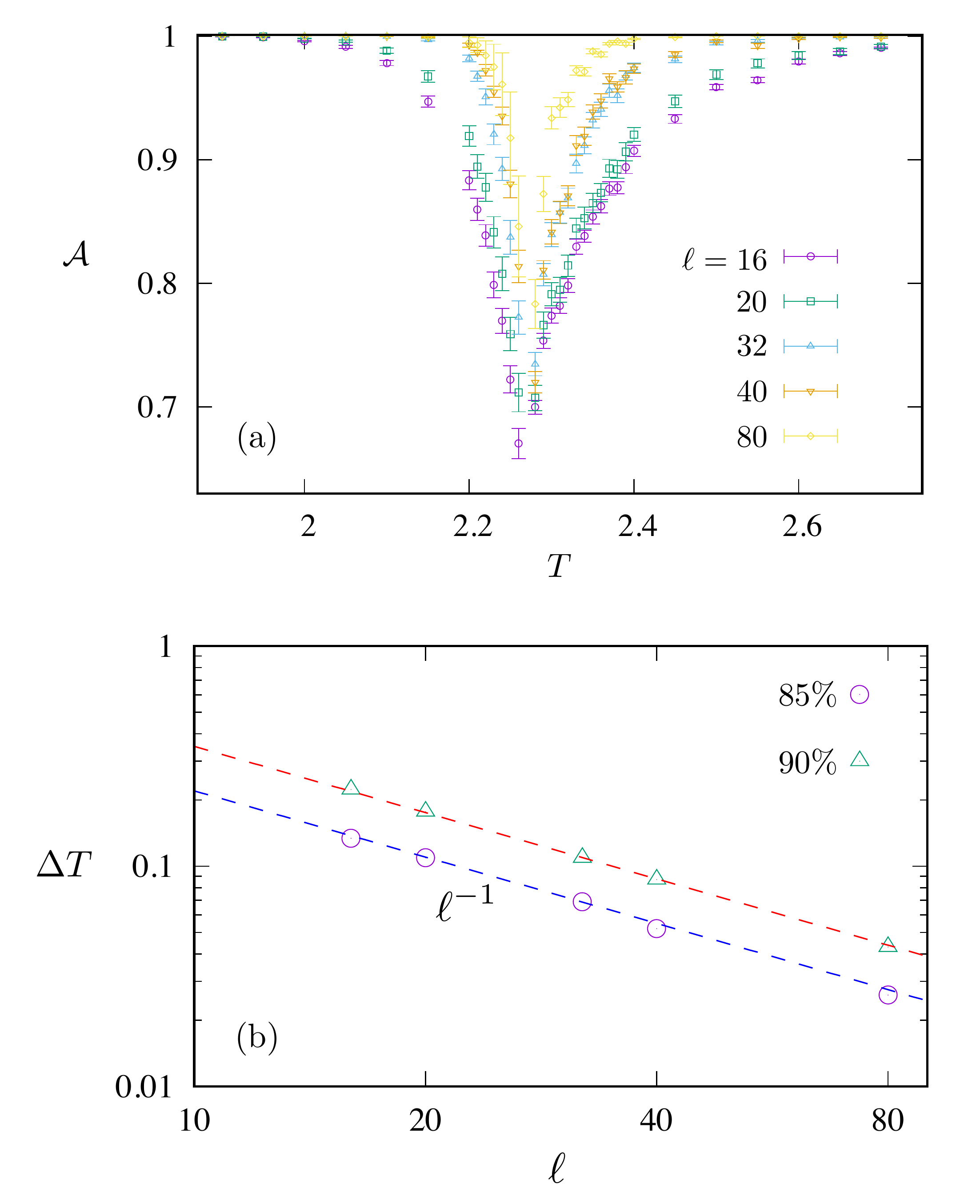}
\caption{(a) Accuracy $\mathcal{A}$ of ML phase classification versus temperature for ML models of different block sizes $\ell$. (b) The temperature window $\Delta T = T^*_+ - T^*_-$ where the classification accuracy $\mathcal{A}$ drops below $85\%$ and $90\%$ versus the block size $\ell$. The dashed lines indicate the power-law behavior $\Delta T \sim \ell^{-1}$.}
\label{fig:prediction_T}  \
\end{figure}

Fig.~\ref{fig:prediction_T}(a) shows the prediction accuracy $\mathcal{A}$ versus temperature for ML models of different block sizes $\ell$. The prediction accuracy here is defined as the ratio of successful predictions over the total number of test data. Overall, as expected, the classification accuracy is better for ML models with larger block size. Moreover, while relatively high classification accuracy can be reached in both the high and low-$T$ phases, especially for ML models with larger $\ell$, the prediction accuracy is significantly reduced in the vicinity of the critical temperature $T_c$ for all ML models. More quantitatively, we define $T^*_\pm$ as the temperatures at which the prediction accuracy drops below a certain threshold; the subscript $\pm$ refers to the high- and low-temperature side of the critical point, respectively. The temperature window $\Delta T = T^*_+ - T^*_-$ within which the accuracy is below threshold $85\%$ and $90\%$ is shown in Fig.~\ref{fig:prediction_T}(b) as a function of ML block size. Interestingly, while this temperature span $\Delta T$ indeed is reduced with increasing block size, we find that its decrease can be very well described by a power-law behavior~$\Delta T \sim \ell^{-1}$.

To understand this result, we note that the divergence of the correlation length as a many-body system approaches the critical point $T_c$ is described by a power-law with an exponent $\nu$:
\begin{eqnarray}
	\xi(T) \sim \frac{1}{|T - T_c|^\nu}.
\end{eqnarray}
For the 2D Ising model, this exponent can be obtained from Onsager's exact solution, which gives $\nu = 1$~\cite{2dising}.
As discussed above, when this correlation length becomes much greater than the size $\ell$ of spin block used for ML model, spin configurations sampled by each block fail to capture the larger spin structure. Consequently, an ML model with input block size~$\ell$ will start to break down when the correlation length reaches the block size 
\begin{eqnarray}
	\xi^* = \xi(T^*) \sim \ell
\end{eqnarray}
Using the power-law relation for $\xi(T)$, one obtains the threshold temperatures on both sides of the critical point
\begin{eqnarray}
	T^*_\pm \sim T_c \pm \mbox{const} \times \ell^{-1/\nu}.
\end{eqnarray}
Defining the temperature span of reduced classification accuracy as $\Delta T = T^*_+ - T^*_-$, we thus obtain the following power-law relation
\begin{eqnarray}
	\label{eq:DT_ell}
	\Delta T \sim \ell^{-1/\nu}.
\end{eqnarray}
For 2D Ising model with $\nu = 1$, the above result is consistent with our numerical simulation shown in Fig.~\ref{fig:prediction_T}(b).

\begin{figure}
\includegraphics[width=0.99\columnwidth]{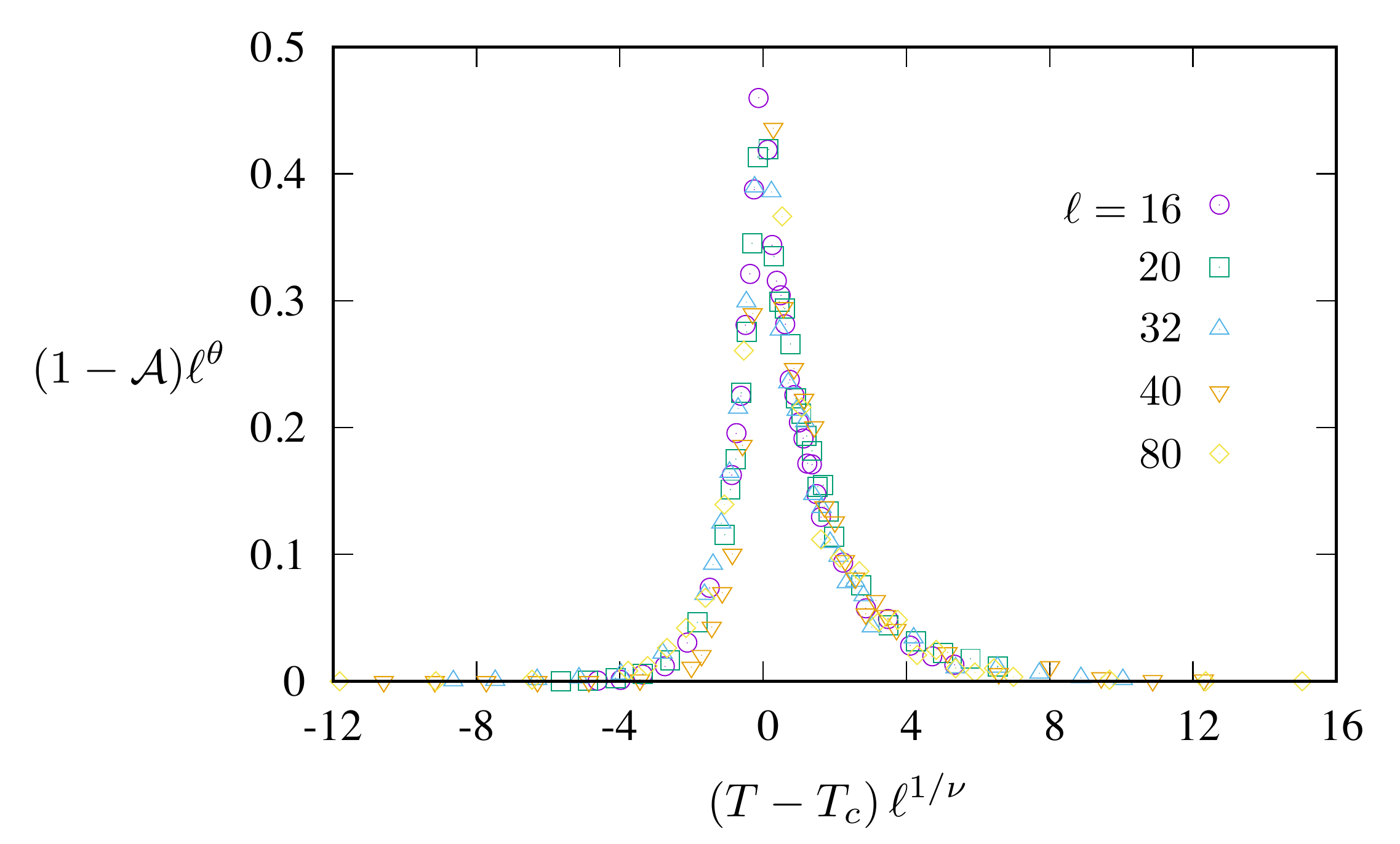}
\caption{Block size scaling analysis of the ML prediction accuracy. The figure shows $(1-\mathcal{A})\cdot \ell^\theta$ versus $(T - T_c ) \ell^\nu$ from prediction results with different block size $\ell$. The exponents $\nu = 1$ is from the 2D Ising model and $\theta = 0.12$. The rather nice data collapsing suggests a relation $ (1 - \mathcal{A}) \cdot \ell^\theta = {F}\bigl( (T-T_c)\cdot \ell^{1/\nu} \bigr)$, where the function ${F}(x)$ is related to the scaling function in Eq.~(\ref{eq:scaling}) as $\Phi(x) = F(x^{1/\nu})$}
\label{fig:scaling}  \
\end{figure}

In fact, the above result suggests a scaling relation for the prediction accuracy similar to the finite-size scaling relations for a continuous phase transition. For example, the order-parameter $m$ of the 2D Ising model exhibits a scaling relation with the system size: $m \sim L^{\beta /\nu} \Psi(L/\xi)$ in the critical regime; here $\Psi(x)$ is a universal function and exponent $\beta = 1/8$ for the 2D Ising model~\cite{2dising}. To apply similar analysis to our case, first the role of system size $L$ is now replaced by the block size $\ell$. Moreover, as the accuracy tends to 1 at both high and low temperatures, this suggests a scaling relation for the prediction error which is defined as the difference $\varepsilon = (1 - \mathcal{A})$. Assuming that the system size $L$ for generating either the training and testing data is much larger than both the correlation length and block size (which is not true when $T \sim T_c$), the prediction error should depend on the ratio $\ell / \xi$ of the only two relevant length scales. This suggests the following scaling relation  
\begin{eqnarray}
	\label{eq:scaling}
	\varepsilon = (1-\mathcal{A}) \sim \ell^{-\theta} \Phi(\ell / \xi), 
\end{eqnarray}
where $\Phi(x)$ is a quasi-universal function which depends on the implementation of ML models, and $\theta$ is an exponent characterizing the improvement of classification accuracy with increasing block size.  This scaling relation is indeed confirmed by our block-size scaling of ML predictions. As shown in Fig.~\ref{fig:scaling}, by plotting $(1-\mathcal{A}) \ell^\theta$ at the $y$-axis and $(\ell / \xi)^{1/\nu} \sim (T - T_c) \, \ell^{1/\nu}$ at the $x$-axis, rather nice data point collapsing was obtained using exponent  $\theta = 0.12$ and $\nu = 1$ from the 2D Ising model. 
The scaling relation Eq.~(\ref{eq:scaling}) shows that, in addition to the rather weak power-law dependence $\ell^{-\theta}$ on the block size, the ML prediction error depends strongly on the ratio of the block size relative to the correlation length. This dependence also underpins the power-law behavior Eq.~(\ref{eq:DT_ell}) for the temperature span with reduced prediction accuracy.  Finally, we note that the scaling relation Eq.~(\ref{eq:scaling}) needs to be modified when the correlation length $\xi$ grows to be greater than the finite system size $L$ used in simulations, which happens in the very vicinity of the critical point. The classification error is expected to depend on the dimensionless ratio $\ell/L$ in this critical regime.




\section{Summary and outlook}

\label{sec:conclusion}

We have presented a scalable ML framework, which is a natural generalization of previous ML-based approaches to structure-property relationships, for the prediction of intensive properties of many-body systems. In particularly, our approach focuses on ML models which can be directly applied to much larger systems without either rebuilding or re-training the model. Specifically, an ML model is developed to produce an estimate of intensive properties based on input of a finite block, characterized by linear size $\ell$, of the system. The training of the ML model is carried out based on dataset of a larger system of linear size $L$ such that the combined prediction of ML model on the $(L/\ell)^d$ blocks is used to optimization parameters of the ML model, where $d$ is the spatial dimension. By including contributions of all blocks of a configuration as a batch for the training, the optimization of the ML model takes into account the structural variations of individual blocks. Application of the trained ML model to large-scale systems is via a random sampling method, i.e. prediction of system-wise intensive properties is obtained by averaging over ML predictions from a large number of randomly selected blocks. 

The two-dimensional Ising model, which is widely used as a benchmark system for ML applications to statistical mechanics and many-body systems, is used to demonstrate our approach. ML models are built for energy-density prediction and phase classification. We show that the breakdown of locality in the vicinity of the critical point leads to uncertainties of ML training, which in turn results in systematic errors for the prediction of energy density. Nonetheless, the systematic errors can be reduced with increasing block size $\ell$ of the ML model.  Furthermore, we show that deterioration in the phase classification accuracy is characterized by the condition $\xi \gtrsim \ell$. The prediction error $\varepsilon$ is found to exhibit a scaling relation which depends strongly on the ratio of block size to the correlation length.

Our proposed ML framework is similar to other ML modeling of structure-property relationship. The crucial difference, due to the demand of scalability, is that the ML model in our approach is to provide a mapping from local structures characterized by linear size $\ell$ to global (intensive) properties of the whole system. The feasibility of this scheme relies on the locality of a many-body system, which is also key to almost all linear-scaling methods. In general, the principle of locality means that local physical quantities only depend on configurations of the immediate surrounding. In our case, on the other hand, the locality assumption means that intensive properties of a system can be inferred from finite-size structures locally. 
Accurate predictions can be achieved only when the linear size $\ell$ of local structures to be modeled by the fixed-size ML model is greater than the structural correlation length~$\xi$ of the system. Interestingly, this means that ML models for accurate phase classification are possible for systems exhibiting first-order phase transitions, since correlation length remains finite during such discontinuous transitions. 

While our scalable ML modeling of the structure-property relationships is based on supervised learning, it is worth noting that phase classification, which is a special case of intensive property prediction, can be achieved using other ML approaches, notably the unsupervised learning methods such as the principle component analysis. Yet, we believe that the scalability of these ML methods are still subject to the fundamental limit due to a diverging correlation length in a continuous phase transition. Detailed analysis of the effects of finite sampling size will be left for future studies. 

For predictions of electronic properties such as energy gap, magnetic susceptibility, or electrical conductivity, although the input to ML models is the directly measurable atomic structures, it should be noted that the relevant correlation length $\xi$ is determined by the electron systems. A representative example is the BP-type ML interatomic potential models for quantum molecular dynamics simulations. In these applications, ML models are designed to predict a local atomic energy based on the neighborhood atomic configurations. Yet, the size of the neighborhood depends on the electronic correlation functions, instead of atomic configurations. As the electron correlation decays exponentially for gapped systems, ML models with large enough $\ell$ are expected to provide an accurate approximation for such systems.

\begin{acknowledgments}
The authors are grateful to useful discussions with Puhan Zhang and Chen Cheng. This work was supported by the US Department of Energy Basic Energy Sciences under Award No. DE-SC0020330. The authors also acknowledge the support of Advanced Research Computing Services at the University of Virginia.
\end{acknowledgments}

\appendix

\section{Descriptor for square-lattice Ising model}

\label{sec:descriptor}

\begin{figure}
\centering
\includegraphics[width=0.9\columnwidth]{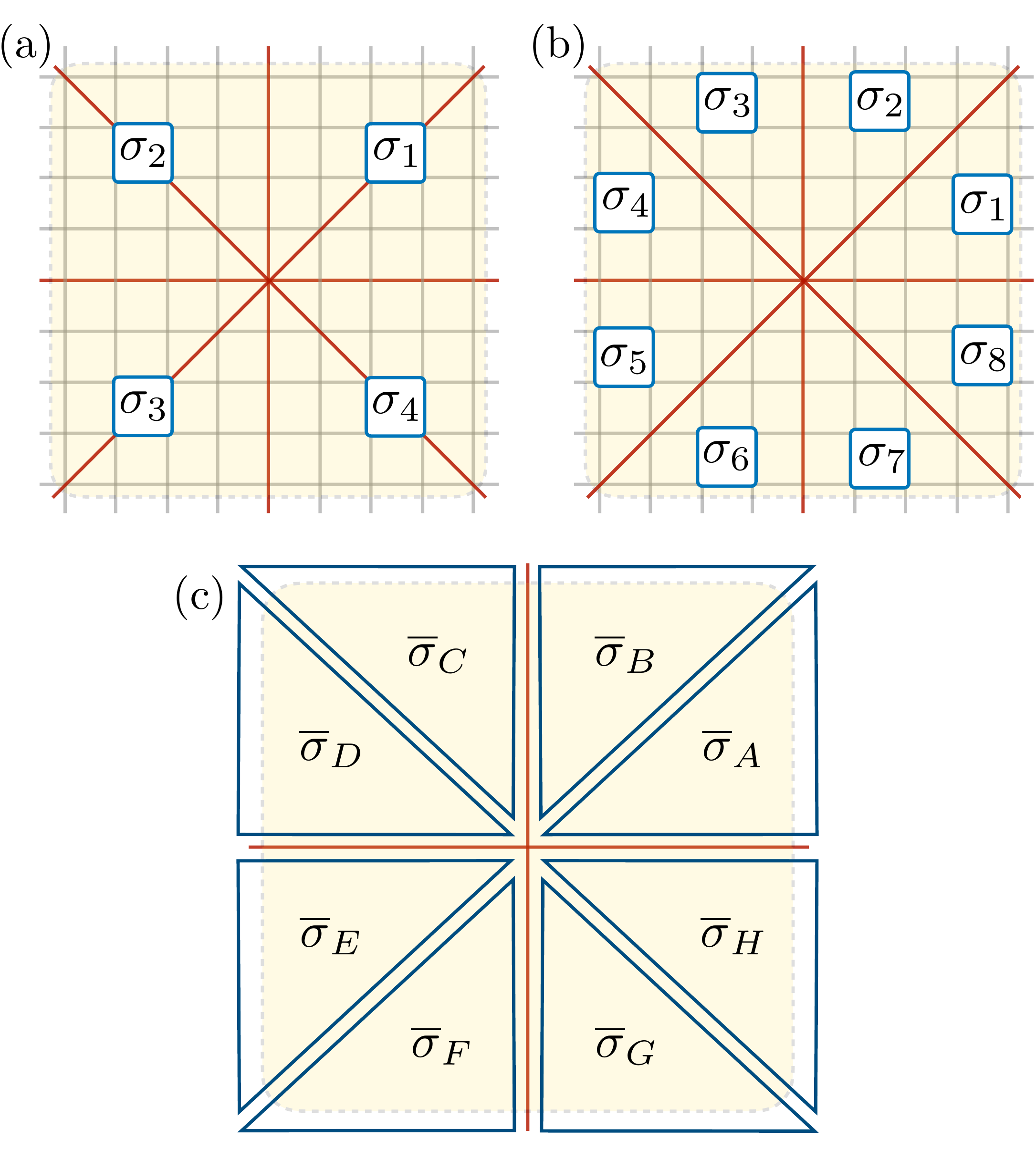}
\caption{Schematic diagrams of (a) 4-sites and (b) 8-sites groups within a spin block. Panel~(c) shows the eight triangular domains used to compute the coefficients of reference IRs.}
\label{fig:descriptor}
\end{figure}

As discussed in Sec.~\ref{sec:framework}, the goal of a descriptor is to preserve lattice symmetry of the original lattice Hamiltonian in the ML model. A general theory and several specific implementations of descriptors in condensed matter systems, especially for lattice models, have recently been presented in Ref.~\cite{zhang22}. In particular, the group-theoretical bispectrum method was generalized to systematically generate feature variables that are invariant under symmetry operations of the on-site point group~\cite{zhang22,ma19}. Here, we apply this method to develop a descriptor for the 2D Ising model. There are two sets of discrete symmetries in the square-lattice Ising model: one is the $Z_2$ symmetry of Ising spins which physically is related to the time-reversal symmetry, and the second is the $D_4$ point-group symmetry of the $\ell \times \ell$ spin block on a square lattice. To account for the $Z_2$ symmetry, one approach is to first build bilinear forms of Ising spins $b_{jk} = \sigma_j \sigma_k$, also known as bond variables, within the spin block. These bilinear variables $b_{jk}$, which are invariant under $Z_2$ transformation $\sigma_i \to -\sigma_i$, are then used as building blocks for the lattice descriptor.

Here, however, we employ a different and simpler approach by constructing invariants of the point-group symmetry first. The $Z_2$ symmetry will be automatically included in the reference method to be discussed below.
To this end, we first note that the Ising spins in an $\ell\times \ell$ block form a high-dimensional reducible representation of the $D_4$ group. For convenience, we use $B_\alpha$ to denote spin configurations in the $\alpha$-th block. The first step of finding invariants under site-symmetry is to decompose $B_\alpha$ into irreducible representations (IRs) of the symmetry group.  This decomposition is considerably simplified due to the lattice geometry. Essentially, since the distance between a neighborhood site-$j$ and the center site-$i$ is invariant under operations of the $D_4$ group, the resultant matrix representation is thus block-diagonalized, with each block corresponding to a group of Ising spins sharing the same distance to the center. In the case of $D_4$, the size of these invariant spin groups is either 4 or 8; see Fig.~\ref{fig:descriptor}(a) and (b). The 4-sites block can be decomposed as: $4 = 1A_1 \oplus 1B_1 \oplus 1E$. The expansion coefficients of each IR are 
\begin{eqnarray}
	& & f_{A_1} = \sigma_a + \sigma_b + \sigma_c + \sigma_d, \nonumber \\
	& & f_{B_1} = \sigma_a - \sigma_b + \sigma_c - \sigma_d, \\
	& & \bm f_E = (\sigma_a - \sigma_c, \, \sigma_b - \sigma_d). \nonumber
\end{eqnarray}
The decomposition of the 8-sites block is: $8 = 1A_1 + 1B_1 + 1A_2 + 1B_2 + 2 E$, with the following coefficients for each IR 
\begin{eqnarray}
	& & f_{A_{1}}  = \sigma_a + \sigma_b + \sigma_c + \sigma_d + \sigma_e + \sigma_f + \sigma_g + \sigma_h, \nonumber \\
	& & f_{A_{2}}  = \sigma_a - \sigma_b + \sigma_c - \sigma_d + \sigma_e - \sigma_f + \sigma_g - \sigma_h, \nonumber \\
	& & f_{B_{1}}  = \sigma_a - \sigma_b - \sigma_c + \sigma_d + \sigma_e - \sigma_f - \sigma_g + \sigma_h, \nonumber \\
	& & f_{B_{2}}  = \sigma_a + \sigma_b - \sigma_c - \sigma_d + \sigma_e + \sigma_f - \sigma_g - \sigma_h, \nonumber \\
	& & \bm f_{(E, 1)}  = (\sigma_a + \sigma_b - \sigma_e - \sigma_f,\  -\sigma_c - \sigma_d + \sigma_g + \sigma_h), \nonumber \\
	& & \bm f_{(E, 2)}  = (\sigma_c - \sigma_d - \sigma_g + \sigma_h,\  \sigma_a - \sigma_b - \sigma_e + \sigma_f).
\end{eqnarray}
As the spin block $B_\alpha$ contains several such 4-sites or 8-sites groups, we expect same IRs appear multiple times in the overall decomposition of the spin block. In the following, we label each IR in the decomposition of $B_\alpha$ as $\Gamma = (\mathtt{T}, r)$, where $\mathtt{T} = A_1, A_2, \cdots$ denotes the symmetry type of the IR, and $r$ indicates different occurrence of the same IR. 
For convenience, we arrange the expansion coefficients of an IR $\Gamma$ into a vector $\bm f_\Gamma = (f_{\Gamma, 1}, f_{\Gamma, 2}, \cdots, f_{\Gamma, n_\Gamma} )$, where $n_\Gamma$ is the dimension of $\Gamma$.

The power spectrum of the representation are given by the amplitudes of the IR coefficients
\begin{eqnarray}
	\label{eq:power}
	p_{\Gamma} = \bigl| \bm f_\Gamma \bigr|^2.
\end{eqnarray}
Since the power spectrum coefficients are obviously invariant under symmetry transformations, they can be used as feature variables for the ML models. However, a descriptor composed only of power spectrum is incomplete since the relative phases between different IRs are ignored. This also means that descriptor contains spurious symmetries as the transformation of each IR is independent of each other without the phase information.  A complete set of feature variables can be obtained from the bispectrum coefficients $b_{\Gamma_1, \Gamma_2, \Gamma_3}$, which are triple products of the expansion coefficients $\bm f_{\Gamma_{1, 2, 3}}$ based on the Clebsch-Gordan coefficients of the point group. Intuitively, they can be viewed as the analog of scalar triple product of 3-dimensional vectors. Not only are the bispectrum coefficients invariant under symmetry transformations, they can also be used to faithfully reconstruct the original disorder configuration~\cite{bartok13,kondor07}.

However, a descriptor based on all the bispectrum coefficients is in fact over-complete as many of them are redundant. Moreover, since the dimension of most IRs of point groups is rather small, the number of bispectrum coefficients is often a very large number, which makes the implementation infeasible. Instead, here we employ the method of reference IR discussed in Ref.~\cite{zhang22} to retain the phase information. The central idea is to first construct an 8-dimensional representation of the spin block $B_\alpha$ based on average of Ising spins over symmetry-related finite regions. As shown in Fig.~\ref{fig:descriptor}(c), an example is given by $(\overline{\sigma}_A, \overline{\sigma}_B, \cdots, \overline{\sigma}_H)$ where $\overline{\sigma}_K = \frac{1}{|K|} \sum_{i \in K} \sigma_i$ is given by the average of all Ising variables  within region-$K$. 

The decomposition of this 8-dimensional representation $\overline{\sigma}_K$ then gives coefficients $ f_{A_1}^*$, $ f^*_{A_2}$, $\cdots$, $\bm f^*_E$ for each symmetry type. These coefficients $\bm f^*_{\mathtt{T}}$ are termed the reference IR coefficients. For each IR, an effective phase can be defined by the following inner product 
\begin{eqnarray}
	\label{eq:phase}
	\eta_\Gamma = \bigl( \bm f_\Gamma \cdot \bm f^*_{\mathtt{T}_\Gamma} \bigr) / \bigl| \bm f_\Gamma \bigr| \bigl| \bm f^*_{\mathtt{T}_\Gamma} \bigr|, 
\end{eqnarray}
where $\mathtt{T}_\Gamma$ is the symmetry type of IR~$\Gamma$. The phase $\eta_\Gamma$, which is an inner product of two IR coefficients, is naturally invariant with respect to symmetry operations. More importantly, by including $\eta_\Gamma$ in the descriptor, the relative phases between different $\bm f_\Gamma$ are now be inferred through the intermediate reference IR coefficients.  Finally, since the IR coefficients $\bm f$ are linear combinations of Ising spins, they acquire a negative sign $\bm f \to -\bm f$ under the $Z_2$ transformation. On the other hand, both the power spectrum Eq.~(\ref{eq:power}) and the relative phases Eq.~(\ref{eq:phase}) are bilinear product of the IR coefficients, the resultant descriptor is automatically invariant with respect to the $Z_2$ symmetry of the Ising model.

\end{document}